\documentclass[aps,prb,twocolumn,showpacs,superscriptaddress]{revtex4}

\usepackage{graphicx}
\usepackage{dcolumn}
\usepackage{bm}
\usepackage{amssymb}
\usepackage{dsfont}

\begin{document}

\title{Quantum correlations in nanostructured two-impurity Kondo systems}

\author{Marco Nizama}
\affiliation{Instituto Balseiro, Centro At\'omico Bariloche, CNEA and CONICET, 8400 Bariloche, Argentina}
\author{Diego Frustaglia}
\email{frustaglia@us.es}
\affiliation{Departamento de F\'{\i}sica Aplicada II, Universidad de Sevilla, E-41012 Sevilla, Spain}
\author{Karen Hallberg}
\affiliation{Instituto Balseiro, Centro At\'omico Bariloche, CNEA and CONICET, 8400 Bariloche, Argentina}

\date{\today}

\begin{abstract}

We study the ground-state entanglement properties of nanostructured Kondo systems consisting of a pair of impurity spins coupled to a background of confined electrons. The competition between the RKKY-like coupling and the Kondo effect determines the development of quantum correlations between the different parts of the system. A key element is the electronic filling due to confinement. An even electronic filling leads to results similar to those found previously for extended systems, where the properties of the reduced impurity-spin subsystem are uniquely determined by the spin correlation function defining a one-dimensional phase space. An odd filling, instead, breaks spin-rotation symmetry unfolding a two-dimensional phase space showing rich entanglement characteristics as, e.g., the requirement of a larger amount of entanglement for the development of non-local correlations between impurity spins. We illustrate these results by numerical simulations of elliptic quantum corrals with magnetic impurities at the foci as a case study.

\end{abstract}


\maketitle


\section{Introduction}

Along the last decades, the emergence of quantum information \cite{NC-00} raised a singular outlook 
in the physical science towards the development of a solid-state quantum engineering.\cite{Z-11} This was supported by the contemporary progress of mesoscopic physics, where the modern laboratory techniques gave a comprehensive access to quantum phenomena in solid-state structures at the nanoscale. 
Presently, many efforts are devoted to the design of nanostructured devices for the controlled 
manipulation of light and matter in the search for functional quantum phenomena. A paradigmatic example is quantum \emph{entanglement}, non-local correlations considered a precious resource for quantum-information processing. 

Among the artificial nanostructures stand out the so-called \emph{quantum corrals}. These are arrays of adatoms deposited on the surface of nobel metals, forming a nearly closed structure within which surface electrons can remain confined.
Quantum corrals have allowed the observation of outstanding effects such as quantum mirages, 
\cite{MLE-Nature00,FH-RMP03} where the introduction of an impurity on one of the foci of an elliptic structure forms a ghost image in the empty one.\cite{HCB-PRL02} 
Several authors have analyzed these experiments with one or more impurities considering different
configurations.\cite{FH-RMP03,agam,aligia,porras,fiete,armandochiappe,lobos,HCB-PRL02,NHA-PRB07,NFH-PB09,kampf,rossi,moon}
It was suggested that two impurities which are located at the foci of the system will
interact strongly as a consequence of the focalizing properties of quantum elliptic corrals. \cite{HCB-PRL02,NHA-PRB07,NFH-PB09,PTP} (Alternative locations apart from the foci were also suggested\cite{WB-PE01} for the production of quantum mirages). This depends on the competition between RKKY and Kondo interactions.\cite{S-PRL06, MJ-PRB09} 
A preliminary study of quantum entanglement in such systems was done in Ref. \onlinecite{PTP}. There, it was found that the localized spins develop a strong mutual entanglement for a small superexchange coupling to the surface electronic states, whereas they disentangle for a strong coupling (when compared to the corresponding  Kondo temperature).

These results suggest that quantum corrals are suitable for essaying possible applications in quantum information and spintronics.\cite{W-Science01} For instance, magnetic impurities added on the foci could be used as quantum units of information (qubits) where quantum correlations (entanglement) mediated by delocalized electronic states are favored by focusing and confinement effects.\cite{NHA-PRB07,NFH-PB09,PTP} Eventually, quantum corrals could serve as elementary prototypes for quantum-information processors provided that logical gates can be implemented by controlled manipulation of the interaction between the magnetic qubits.\cite{MS-PRL04} 

Here we study the entanglement developed between two magnetic impurities embedded in a confined electronic environment. This corresponds to a nanostructured Kondo system, where localized  magnetic impurities are subject to an effective (RKKY-like \cite{RKKY}) exchange interaction mediated by electron-in-a-box states coupled antiferromagnetically (AFM) to the impurities. The key element in our discussion is the electron confinement.\cite{note-2} Here, in contrast to extended systems with electrons organized in a Fermi sea,\cite{CMc-PRA06}   
the strength of the quantum correlations is determined by the electron \emph{ filling}. For an extended Kondo model, the ground state of the composed two-impurity/electron system forms a spin singlet. \cite{JVW-PRL88} Cho and McKenzie \cite{CMc-PRA06} have shown that this reduces the two-impurity subsystem to a (mixed) rotationally invariant Werner state.\cite{W-PRA89} This is a particular family of (mainly mixed) states characterized by a single parameter identified with the impurity-spin correlation function, defining a one-dimensional phase space running continuously from a fully uncorrelated classical ensamble to a fully entangled (singlet) state. We shall see that confined Kondo systems with even electronic filling share similar properties. An odd electronic filling, instead, takes the two-impurity subsystem away from the Werner-state family to a region of the Hilbert space with broken rotational symmetry and richer entanglement characteristics. Such an extended family cannot be fully described by a single observable. Instead, as we shall see, it unfolds a two-dimensional phase space. 

Moreover, it is known that entanglement does not guarantee \emph{non-locality}: while the former refers to the (non)separability of a quantum state, the latter is defined only by the violation of Bell-like inequalities (meaning that quantum correlations cannot be modeled by a hidden classical-variable theory). Within the Werner-state family, indeed, there exists a subclass of entangled states which do not violate Bell's inequalities. The development of non-locality requires to overcome a minimal amount of entanglement (measured in terms of, e.g., the \emph{concurrence} \cite{W-PRL98}). We shall see that an odd electronic filling generally constrains the impurity-spin system to develop a larger amount of entanglement for the violation of Bell's inequalities when compared to the case of even electronic filling. 

The work is organized as follows. 
In Sec.~\ref{IESISS} we discuss the role played by the electronic filling on the reduced impurity-spin system based on symmetry arguments, only. In Sec.~\ref{entan} we introduce a phase-space diagram classifying the entanglement properties of the impurity-spin system and discuss the 
conditions necessary for the development of non-local correlations. Furthermore, in Sec.~\ref{EISCE} we study  the complementary entanglement arising between the impurity-spins and a confined electronic bath.  
Finally, in Sec.~\ref{NSQC} we present numerical results corresponding to Kondo-model simulations for an elliptic quantum corral discussing the distribution of solutions over the entanglement phase-space diagram.


\section{The influence of the electronic states on the impurity spin system}
\label{IESISS}

\subsection{Full (pure) electron/impurity state}

\begin{figure}[t!]
\includegraphics[width=0.95 \columnwidth,clip]{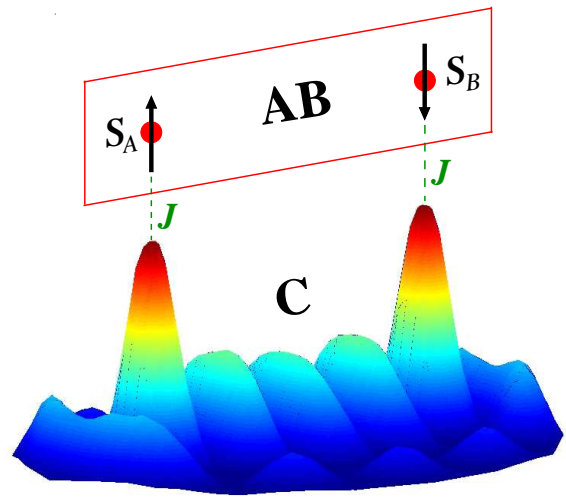}
\caption{
Nanostructured Kondo system: The figure depicts two magnetic impurities $A$ and $B$ on the foci  
of an elliptic quantum corral. Confined electronic states $C$ of given filling couple antiferromagnetically to the impurity spins, developing either local Kondo states or RKKY-like superexchange between impurity-spins depending on the coupling strength $J$. 
} 
\label{fig-ks}
\end{figure}

We start by considering a generic pure state $|\Psi\rangle $ of the composed system $ABC$ with a definite $z$-projection of the total spin $S_{\rm T}^z$, where $A$ and $B$ name the impurity spins and $C$ identifies a set of $N$ electrons confined within an arbitrary nanoscopic system (see  Fig.~\ref{fig-ks} for the particular case of a quantum corral):
\begin{equation}
|\Psi\rangle = 
a_1 |\uparrow \uparrow \rangle |\Phi_1\rangle + 
a_2 |\uparrow \downarrow \rangle |\Phi_2\rangle + 
a_3 |\downarrow \uparrow \rangle |\Phi_3\rangle + 
a_4 |\downarrow \downarrow \rangle |\Phi_4\rangle.  
\label{Psi}
\end{equation}
Here, the first ket of each term represents the two-impurity spin subsystem while the 
$|\Phi_n\rangle$ are $N$-particle electronic states of the form
\begin{eqnarray}
|\Phi_1\rangle =\sum_n \phi_n^1 |\gamma_n^-\rangle,\\
|\Phi_2\rangle =\sum_n \phi_n^2 |\gamma_n^0\rangle,\\
|\Phi_3\rangle =\sum_n \phi_n^3 |\gamma_n^0\rangle,\\
|\Phi_4\rangle =\sum_n \phi_n^4 |\gamma_n^+\rangle.
\end{eqnarray}
The sets $\{ |\gamma_n^-\rangle\}$, $\{ |\gamma_n^0\rangle\}$, and $\{ |\gamma_n^+\rangle\}$ are orthonormal bases of $N$-electron states with $z$-projection of the spin $S_{\rm e}^z=S_{\rm T}^z-1$, $S_{\rm T}^z$, and $S_{\rm T}^z+1$, respectively, where $S_{\rm T}^z$ (total $z$-component of the spin, including impurities) has a definite value. The electronic states $|\Phi_n\rangle$ are such that $\langle\Phi_1|\Phi_n\rangle=\delta_{1n}$ and $\langle\Phi_4|\Phi_n\rangle=\delta_{4n}$, while $\langle\Phi_2|\Phi_3\rangle \neq 0$ in general since both states belong to the same spin subspace. Without loss of generality, in Eq.~(\ref{Psi}) we assume real and positive probability amplitudes $a_n$ such that $\sum_n a_n^2=1$. 

The symmetry properties of $|\Psi\rangle$ upon spin rotation depend on the electronic filling $N$, a constraint
for the total spin. For an even electron number, the {\it ground state} of the antiferromagnetic Kondo model for the composed $ABC$ system is a spin singlet, invariant under joint spin rotation. \cite{CMc-PRA06,JVW-PRL88} The situation changes for odd electron number, where total spin is a half-integer. This is reflected in the reduced two-impurity spin system $AB$ with striking consequences on quantum correlations, as we shall discuss.

%
%

\subsection{Reduced two-impurity spin density matrix}

The reduced (mixed) $4 \times 4$ density matrix for the localized spins $\rho_{AB}$ is obtained by tracing out the electronic degrees of freedom $C$ from the full (pure) density matrix $\rho=|\Psi\rangle \langle \Psi|$.  In the ordered basis $\{ |\uparrow \uparrow \rangle, |\uparrow \downarrow \rangle, |\downarrow \uparrow \rangle, |\downarrow \downarrow \rangle \}$ that yields
\begin{equation}
\rho_{AB}=\left( \begin{array}{cccc}
a_1^2 & 0 & 0 & 0 \\
0 & a_2^2 & a_2 a_3 \langle\Phi_3|\Phi_2\rangle& 0 \\
0 & a_2 a_3 \langle\Phi_3|\Phi_2\rangle^* & a_3^2 & 0\\
0 & 0 & 0 & a_4^2
\end{array} \right). 
\label{rho-1}
\end{equation}
We first notice that $\rho_{AB}$ does not mix impurity states with different spin projection. This results from assuming that both the electron number and the projection of the total spin along the $z$ axis are good quantum numbers. Further consequences follow from considering those symmetry constraints present  in (though not limited to) elliptic quantum corrals with magnetic impurities at the foci. The reflection symmetry along the minor axis implies that $\rho_{AB}$ must be invariant under magnetic impurity interchange. This means $\rho_{AB}=\rho_{BA}$, with $a_2=a_3$ and $\langle\Phi_3|\Phi_2\rangle = \langle\Phi_3|\Phi_2\rangle^*\equiv \cos \varphi$, where the phase $0 \le \varphi \le \pi$ parametrizes the overlap between $|\Phi_2 \rangle$ and $|\Phi_3\rangle$. Under such symmetry constraints, Eq.~(\ref{rho-1}) reduces to 
\begin{equation}
\rho_{AB}=\left( \begin{array}{cccc}
1-2a_2^2-a_4^2 & 0 & 0 & 0 \\
0 & a_2^2 & a_2^2 \cos \varphi & 0 \\
0 & a_2^2 \cos \varphi & a_2^2 & 0\\
0 & 0 & 0 & a_4^2
\end{array} \right), 
\label{rho-2}
\end{equation}
with $0 \le a_1^2= 1-2a_2^2-a_4^2 \le 1$, $0 \le a_2^2 \le 1/2$ and $0 \le a_4^2 \le 1$.
Notice that orthogonal electronic components $|\Phi_2 \rangle$ and $|\Phi_3\rangle$ ($\cos \varphi=0$) would lead to vanishing coherence (off-diagonal) terms in $\rho_{AB}$. This would prevent the impurity spins from developing mutual quantum correlations such as entanglement (see Sec.~\ref{entan} for a detailed discussion). 

By observing the central $4\times4$ block of matrix (\ref{rho-2}), we notice that $\rho_{AB}$ can be expanded as 
\begin{eqnarray}
\label{rho-st}
\rho_{AB} &=& 
(1-2a_2^2-a_4^2) |\uparrow \uparrow \rangle \langle \uparrow \uparrow| +
a_4^2 |\downarrow \downarrow \rangle \langle\downarrow \downarrow| \\
&+&\frac{a_2^2}{2} \sum_{\delta=\pm 1}(1+ \delta \cos\varphi) |\psi^\delta \rangle\langle\psi^\delta | \nonumber,  
\nonumber
\end{eqnarray}
where $|\psi^{\pm}\rangle=(|\uparrow \downarrow \rangle \pm | \downarrow \uparrow\rangle)/\sqrt{2}$ are fully entangled triplet ($+$) and singlet ($-$) Bell states. Interestingly, pure triplet and singlet states are  possible for $a_2^2=1/2$ (implying $a_1=a_4=0$) provided that $\cos \varphi = \pm 1$, respectively.

\subsection{Hilbert-Schmidt decomposition and Werner-state component}
\label{HSWS}

It is helpful to rewrite Eq.~(\ref{rho-2}) by introducing the Hilbert-Schmidt decomposition \cite{NC-00}
\begin{equation}
\rho_{AB} =\frac{1}{4} \sum_{\alpha,\beta=0,x,y,z} r_{\alpha \beta}~\sigma_A^\alpha \otimes \sigma_B^\beta
\label{rho-3}
\end{equation}
with physical observables 
\begin{equation}
r_{\alpha \beta}=\text{Tr}(\sigma_A^\alpha \sigma_B^\beta \rho_{AB}) \in \mathds{R},
\end{equation}
where $\sigma_j^0 \equiv \mathds{1}_j$ and $\sigma_j^\nu$ are the $2 \times 2$ identity and Pauli matrices, respectively ($j=A,B$ and $\nu=x,y,z$). By assuming invariance under magnetic impurity interchange and spin rotation symmetry around the $z$ axis, Eq.~(\ref{rho-3}) reduces to 
\begin{widetext}
\begin{equation}
\rho_{AB}=\frac{1}{4}\left( \begin{array}{cccc}
1+r_{zz} + 2 r_{0z} & 0 & 0 & 0 \\
0 & 1- r_{zz} & 2 r_{xx} & 0 \\
0 & 2 r_{xx} & 1- r_{zz} & 0\\
0 & 0 & 0 & 1+r_{zz} - 2 r_{0z}
\end{array} \right), 
\label{rho-4}
\end{equation}
\end{widetext}
where we find $r_{xx}=r_{yy}=2 a_2^2 \cos \varphi$, $r_{zz}=1-4 a_2^2$, and 
$r_{0z}=1- 2 (a_2^2+a_4^2)$ by direct comparison with Eq.~(\ref{rho-2}). 

When the interaction between electron and impurity spins is antiferromagnetic and isotropic, the {\it ground state} of the composed $ABC$ system, $|\Psi_{\rm G}\rangle$, is either a spin singlet or a doublet according to the electronic filling (even or odd, respectively). In any of these cases the reduced density matrix of Eq.~(\ref{rho-4}) satisfies $r_{xx}=r_{zz}= r$ \cite{note} (an $r_{xx} \neq r_{zz}$, instead, is still possible for {\it excited states}). Under such conditions we can conveniently write
\begin{equation}
\rho_{AB}=\rho_{AB}^{\rm W}+\Lambda_{AB}
\label{rho-5}
\end{equation}
with
\begin{eqnarray}
\rho_{AB}^{\rm W}&=&\frac{1}{4} \left (  \mathds{1}_{4 \times 4} + r \sum_{\alpha=x,y,z} \sigma_A^\alpha \otimes \sigma_B^\alpha \right), \\
\Lambda_{AB}&=&\frac{r_{0z}}{4} \left (  \sigma_A^z \otimes \mathds{1}_B + 
\mathds{1}_A \otimes \sigma_B^z \right ).
\label{rho-z}
\end{eqnarray}
Here, $\rho_{AB}^{\rm W}$ is a Werner-state density matrix.\cite{CMc-PRA06,W-PRA89}  This is most commonly written as
\begin{equation}
\rho_{AB}^{\rm W}=\frac{1-p_{\rm s}}{3} \mathds{1}_{4 \times 4} + \frac{4p_{\rm s}-1}{3} |\Psi^-\rangle\langle  \Psi^-|,
\end{equation}
parametrized by the so-called singlet fidelity $p_{\rm s}\equiv \langle  \Psi^-|\rho_{AB}^{\rm W}|\Psi^-\rangle=(1-3r)/4$ with $0\le p_{\rm s}\le1$. Remarkably, $\rho_{AB}^{\rm W}$ is determined by one single observable: the spin-spin correlation function $\langle {\bf S}_A \cdot {\bf S}_B\rangle=3r/4=1/4-p_{\rm s}$, with $-3/4 \le \langle {\bf S}_A \cdot {\bf S}_B\rangle \le 1/4$.
The $\Lambda_{AB}$ of Eq.~(\ref{rho-z}), instead, is a traceless contribution to $\rho_{AB}$ (namely, it is not a density matrix itself) determined {\it only} by the $z$-projection of the impurity spin, since $r_{0z}= \langle S^z_A + S^z_B\rangle$. The $\rho_{AB}$ is then fully determined by two observables, only: $\langle {\bf S}_A \cdot {\bf S}_B\rangle$ and $\langle S^z_A + S^z_B\rangle$. This is a consequence of the setting $r_{xx}=r_{zz}$.  Otherwise, $\langle {\bf S}_A \cdot {\bf S}_B\rangle$ and $\langle S^z_A + S^z_B\rangle$ would not be sufficient to determine $\rho_{AB}$ completely, providing only a {\it partial} tomography of the quantum state. For $\langle S^z_A + S^z_B\rangle=0$, $\rho_{AB}$ reduces to a Werner state. This is the case when the full (impurity/electron) state is a singlet, as discussed in Ref.~\onlinecite{CMc-PRA06}. A $\langle S^z_A + S^z_B\rangle \neq 0$ takes $\rho_{AB}$ away from the Werner-state subspace modifying the entanglement features (see Sec.~\ref{entan} for a detailed discussion).

The setting $r_{xx}=r_{zz}$ also introduces some constraints on the electronic parameters of Eq.~(\ref{rho-2}). By noticing that $r_{xx}=2 a_2^2 \cos \varphi$ and $r_{zz}=1-4a_2^2$, we find $\cos \varphi= -2+1/2a_2^2$ with $1/6 \le a_2^2 \le 1/2$. This allows the development of pure singlets $|\psi^-\rangle$ for $a_2^2=1/2$ at the same time it forbids pure triplets $|\psi^+\rangle$ (see Eq.~(\ref{rho-st})). For $a_2^2=1/4$, instead, all coherence terms in Eq.~(\ref{rho-2}) vanish together with $\cos \varphi$.

\subsection{Relation between $\langle {\bf S}_A \cdot {\bf S}_B\rangle$ and $\langle S^z_A + S^z_B\rangle$}

While the observables $\langle {\bf S}_A \cdot {\bf S}_B\rangle$ and $\langle S^z_A + S^z_B\rangle$ are in principle independent, the introduction of symmetry constraints can restrict such freedom. However, symmetry alone is not sufficient for determining the ultimate relation between them. This, as expected, is eventually given by the full Hamiltonian. 

From Eqs.~(\ref{rho-2}) and (\ref{rho-4}) we find $\langle {\bf S}_A \cdot {\bf S}_B\rangle=(2 r_{xx}+r_{zz})/4=1/4+(\cos \varphi -1) a_2^2$ and $\langle S^z_A + S^z_B\rangle = r_{0z} = 1- 2 (a_2^2+a_4^2)$. The condition $r_{xx}=r_{zz}$ yields 
\begin{eqnarray}
\label{SzFs}
\langle S^z_A + S^z_B\rangle
&=&\frac{1}{2}+\frac{2}{3} \langle {\bf S}_A \cdot {\bf S}_B\rangle -2a_4^2 \\
&=& -\frac{1}{2}-\frac{2}{3} \langle {\bf S}_A \cdot {\bf S}_B\rangle +2a_1^2.
\nonumber
\end{eqnarray}
This equation is obtained from the application of symmetry constraints, only. From it we can identify several scenarios. For the symmetric case $a_1^2=a_4^2$ we find $\langle S^z_A + S^z_B\rangle=0$ while $\langle {\bf S}_A \cdot {\bf S}_B\rangle$ runs freely from $-3/4$ to $1/4$. This corresponds to the Werner state $\rho_{AB}=\rho_{AB}^{\rm W}$ discussed in Ref.~\onlinecite{CMc-PRA06} (see Eq.~(\ref{rho-5})).  We further see that when either $a_1^2$ or $a_4^2$ equals zero the relation between $\langle {\bf S}_A \cdot {\bf S}_B\rangle$ and $\langle S^z_A + S^z_B\rangle$ must be {\it linear}.  Otherwise, they are free to develop a more general {\it non-linear} behavior which must be determined by solving the Hamiltonian equation. We illustrate this in the following section.


\section{Entanglement in the reduced spin system}
 \label{entan}
 
\subsection{Concurrence and phase-space representation}
 
A standard measure of the entanglement between two $1/2$ spins or qubits is the concurrence ${\cal C}$.\cite{W-PRL98} This runs from zero for separable (disentangled) states to one for maximally entangled (Bell) states.  For the reduced two-impurity spin density matrix $\rho_{AB}$ of Eq.~(\ref{rho-4}) we find a concurrence 
\begin{equation}
\label{C-1}
{\cal C}(\rho_{AB})
=\text{max} \left \{  |r_{xx}|- \frac {1}{2} \sqrt{(1+r_{zz})^2-4r_{0z}^2},~0 \right \}.
\end{equation}
This can be fully determined by $\langle {\bf S}_A \cdot {\bf S}_B\rangle$ and $\langle S^z_A + S^z_B\rangle$ provided that $r_{xx}=r_{zz}$, reducing to
\begin{widetext}
\begin{equation}
{\cal C}(\rho_{AB})=\text{max} \left \{  \frac {4}{3} |\langle {\bf S}_A \cdot {\bf S}_B\rangle|- \frac {1}{2} \sqrt{\left (\frac{4}{3} \langle {\bf S}_A \cdot {\bf S}_B\rangle+1\right )^2- 4\langle S^z_A + S^z_B\rangle^2},~0     \right \}.
\label{C-2}
\end{equation}
\end{widetext}

\begin{figure}[t!]
\includegraphics[width=0.95 \columnwidth,clip]{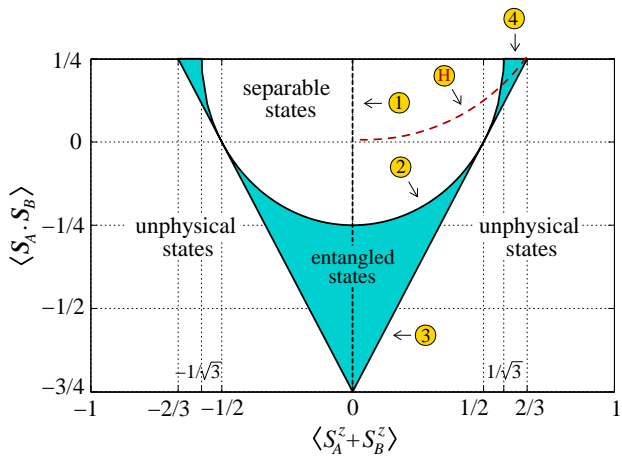}
\caption{
Phase-space diagram for the reduced impurity-spin system AB: The figure classifies the entanglement properties of $\rho_{AB}$ according to Eq.~(\ref{C-2}), parametrized by the observables $\langle S^z_A + S^z_B\rangle$ and $\langle {\bf S}_A \cdot {\bf S}_B\rangle$. 
Curve 1 along $\langle S^z_A + S^z_B\rangle=0$ corresponds to Werner states. Curve 2 defines the critical limit separating entangled from disentangled states, along which the concurrence ${\cal C}$ vanishes. Curve 3 marks the limit for physically possible states compatible with the symmetry constraints.   Curve 4 indicates a subspace of entangled, mixed triplet states. Curve H corresponds to numerical simulations on an elliptic quantum corral for an odd filling. 
} 
\label{fig-psd}
\end{figure}

In Fig.~\ref{fig-psd} we depict a phase-space diagram classifying the entanglement properties of the $\rho_{AB}$ given in Eq.~(\ref{rho-5}) as a function of $\langle {\bf S}_A \cdot {\bf S}_B\rangle$ and $\langle S^z_A + S^z_B\rangle$.
The shaded area corresponds to the entangled phase with ${\cal C}>0$, according to Eq.~({\ref{C-2}}). This region is defined between curve 2: $\langle {\bf S}_A \cdot {\bf S}_B\rangle=-1/4+\left(1-\sqrt{1-3\langle S^z_A + S^z_B\rangle^2}~\right)/2$, and curve 3: $\langle {\bf S}_A \cdot {\bf S}_B\rangle = -3/4 + (3/2) |\langle S^z_A + S^z_B\rangle|$, along which ${\cal C}=0$ and ${\cal C}=\left|2|\langle S^z_A + S^z_B\rangle|-1\right|$, respectively. The region defined above curve 2: $\langle {\bf S}_A \cdot {\bf S}_B\rangle \ge -1/4+\left(1-\sqrt{1-3\langle S^z_A + S^z_B\rangle^2}~\right)/2$, encloses the subspace of separable states (${\cal C}=0$). The region below curve 3: $\langle {\bf S}_A \cdot {\bf S}_B\rangle < -3/4 + (3/2) |\langle S^z_A + S^z_B\rangle|$, instead, corresponds to unphysical states with imaginary concurrence.\cite{note-1} 
For $\langle S^z_A + S^z_B\rangle=0$, Eq.~(\ref{C-2}) reduces to the well-known concurrence for Werner states ${\cal C}(\rho_{AB}^{\rm W})=\text{max}\{ 2p_{\rm s}-1=-(1/2+2\langle {\bf S}_A \cdot {\bf S}_B\rangle),~0\}$ (see Sec.~\ref{HSWS}). This family of states is represented in the phase-space diagram of Fig.~\ref{fig-psd} by curve 1 (dashed line), where the value $\langle {\bf S}_A \cdot {\bf S}_B\rangle_{\rm c}=-1/4$ ($p_s^{\rm c}=1/2$) defines a critical point separating entangled from product Werner states. 

Maximum entanglement (${\cal C}=1$) is reached only at a single point ($\langle S^z_A + S^z_B\rangle=0$, $\langle {\bf S}_A \cdot {\bf S}_B\rangle=-3/4$) corresponding to the singlet $|\psi^-\rangle$, from which ${\cal C}$ decreases monotonously (eventually vanishing) as $\langle {\bf S}_A \cdot {\bf S}_B\rangle$  approaches zero. From there on, ${\cal C}$ increases again up to ${\cal C}=1/3-\sqrt{4/9-\langle S^z_A + S^z_B\rangle^2}$ for $\langle {\bf S}_A \cdot {\bf S}_B\rangle=1/4$ and $1/\sqrt{3} \le |\langle S^z_A + S^z_B\rangle| \le 2/3$ (segment 4 in Fig.~\ref{fig-psd}), corresponding to mixed triplet states (i.e., incoherent superpositions of $|\psi^+\rangle, |\uparrow \uparrow \rangle$, and $|\downarrow \downarrow \rangle$). We then conclude that, remarkably, a $\langle S^z_A + S^z_B\rangle \neq 0$ allows for entanglement {\it beyond the limit imposed to Werner states}, namely, ${\cal C}>0$ for $\langle {\bf S}_A \cdot {\bf S}_B\rangle > -1/4$.

%
%

\subsection{Violation of Bell inequalities}

The presence of quantum entanglement does not guarantee the development of the non-local quantum correlations present in maximally entangled (Bell) states as, e.g., the spin singlet. Non-locality is properly  identified through the (violation of) Bell-CHSH (Clauser-Horne-Shimony-Holt) inequalities.\cite{CHSH-PRL69} Some entangled (non-separable) states can certainly satisfy Bell-CHSH inequalities: a well-known example are the Werner states with $-3/4\sqrt{2} \le \langle {\bf S}_A \cdot {\bf S}_B\rangle < -1/4$ ($1/2<p_{\rm s} \le (1+3/\sqrt{2})/4$). \cite{HHH-PLA95} This means that, in such states, the statistical properties ascribed to the entanglement can be reproduced by using a hidden variable model, i.e., quantum features can be modeled classically. 

\begin{figure}[t!]
\includegraphics[width=0.95 \columnwidth,clip]{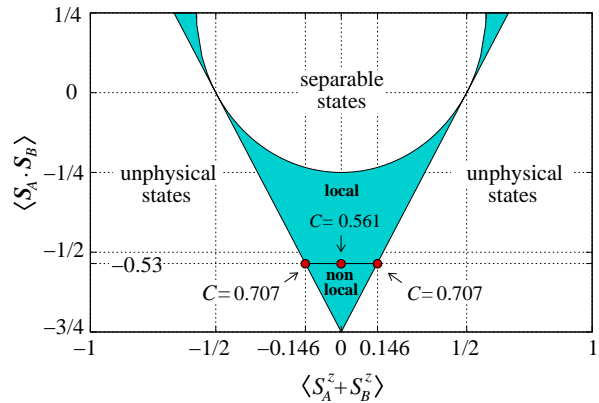}
\caption{
Locality vs. non-locality in the reduced impurity-spin system: The value $\langle {\bf S}_A \cdot {\bf S}_B\rangle=-3/4\sqrt{2}\approx-0.53$ determines a critical limit separating entangled states subject to non-local correlations (violating Bell-like inequalities) from those compatible with a local hidden-variable model (satisfying Bell-like inequalities). The minimal amount of entanglement required for the development of non-local correlations increases with $|\langle S^z_A + S^z_B\rangle|$.
} 
\label{fig-vb}
\end{figure}

Following the criteria introduced in Ref.~\onlinecite{HHH-PLA95}, we find that the $\rho_{AB}$ of Eq.~(\ref{rho-5}) violates the corresponding Bell-CHSH inequality for $\langle {\bf S}_A \cdot {\bf S}_B\rangle < -3/4\sqrt{2}$. Surprisingly, this upper bound coincides with the result found for Werner states. However, the presence of a finite $\langle S^z_A + S^z_B\rangle$ has some consequences on the minimal amount of entanglement (measured in terms of concurrence) needed for violating the inequality. By setting $\langle {\bf S}_A \cdot {\bf S}_B\rangle=-3/4\sqrt{2}\approx-0.53$ in Eq.~(\ref{C-2}), we find 
\begin{equation}
{\cal C}|_{\langle {\bf S}_A \cdot {\bf S}_B\rangle=\frac{-3}{4\sqrt{2}}}=\frac{1}{\sqrt{2}}-\frac{1}{2}\sqrt{\left(1-\frac{1}{\sqrt{2}}\right)^2-4 \langle S^z_A + S^z_B\rangle^2}.
\end{equation}
Here we see that the minimal concurrence required for violating the Bell-CHSH inequality runs from ${\cal C}=3/2\sqrt{2}-1/2 \sim 0.561$ for Werner states ($\langle S^z_A + S^z_B\rangle=0$) to ${\cal C}=1/\sqrt{2}\approx 0.707$ ($|\langle S^z_A + S^z_B\rangle|=(1-1/\sqrt{2})/2 \approx 0.146$), see Fig.~\ref{fig-vb}. This means that  
states with a finite $\langle S^z_A + S^z_B\rangle$ {\it require a larger amount of entanglement for violating locality} than Werner states.

%
%


\section{Entanglement between impurity spins and confined electron}
\label{EISCE}

\begin{figure}[t!]
\includegraphics[width=0.95 \columnwidth,clip]{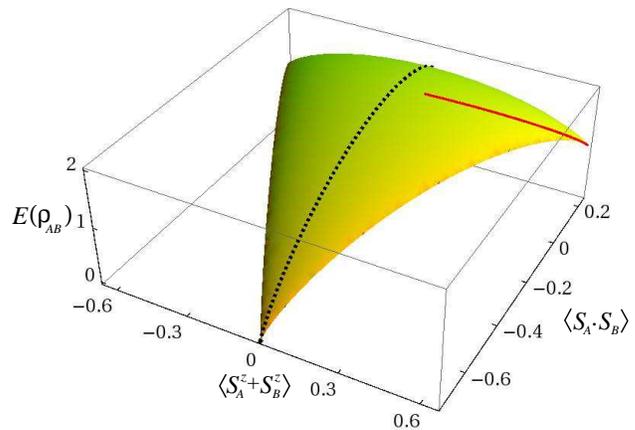}
\caption{
Entanglement between confined electrons and impurity spins given by the von Neumann entropy $E(\rho_{AB})$ of Eq.~(\ref{E-AB}) parametrized 
by the observables $\langle S^z_A + S^z_B\rangle$ and $\langle {\bf S}_A \cdot {\bf S}_B\rangle$. 
Broken-line and solid curves correspond to numerical simulations on an elliptic quantum corral for even and odd electrons respectively.
} 
\label{fig-E}
\end{figure}

Entanglement in a quantum system depends on the partition choice, namely, on the particular
(pair of) subsystems under study. We now explore the entanglement developed between the 
impurity-spins and the electronic subsystems. A good measure of this entanglement is the von 
Neumann entropy $E$ of the reduced impurity-spin density matrix $\rho_{AB}$,
\cite{BBPS-PRA96,NC-00} defined as
\begin{equation}
\label{E-AB}
E(\rho_{AB})=-\sum_{n=1}^4 \omega_n \log_2\omega_n,
\end{equation}
where  
\begin{eqnarray}
\omega_1&=&\frac{1}{4}+\frac{1}{3} \langle {\bf S}_A \cdot {\bf S}_B\rangle + \frac{1}{2} \langle S^z_A + S^z_B\rangle, \nonumber \\
\omega_2&=&\frac{1}{4}+\frac{1}{3} \langle {\bf S}_A \cdot {\bf S}_B\rangle - \frac{1}{2} \langle S^z_A + S^z_B\rangle,  \nonumber  \\
\omega_3&=&\frac{1}{4}-\frac{1}{3} \langle {\bf S}_A \cdot {\bf S}_B\rangle + \frac{2}{3} |\langle {\bf S}_A \cdot {\bf S}_B\rangle|,  \nonumber \\
\omega_4&=&\frac{1}{4}-\frac{1}{3} \langle {\bf S}_A \cdot {\bf S}_B\rangle - \frac{2}{3} |\langle {\bf S}_A \cdot {\bf S}_B\rangle|,  \nonumber 
\end{eqnarray}
are the eigenvalues of $\rho_{AB}$. The $E(\rho_{AB})$ runs from zero for separable states to two for maximally entangled states. Absolute maxima and minima of $E(\rho_{AB})$ are found along the curve defined by $\langle S^z_A + S^z_B\rangle = 0$ at points $\langle {\bf S}_A \cdot {\bf S}_B\rangle = 0$ and $\langle {\bf S}_A \cdot {\bf S}_B\rangle = -3/4$, respectively (see Fig.~\ref{fig-E}). This coincides with what was found in the case of global impurity-spin/electron singlet states,\cite{CMc-PRA06} where $E(\rho_{AB})=2$ indicates the formation of local Kondo singlets between impurity spins and electrons. 
$E(\rho_{AB})=0$ corresponds to the development of an impurity-spin singlet separated from the electronic system. The introduction of a finite $\langle S^z_A + S^z_B\rangle$ produces a decorrelation between impurity-spins and electrons, reducing the amount of entanglement $E(\rho_{AB})$ as shown in Fig.~\ref{fig-E}. Such decorrelation, however, is not complete (except in the vicinity of the impurity-spin singlet), so that we can generally claim that the impurity-spin subsystem shows a relatively high correlation with the electronic degrees of freedom. 

Complementary information can be obtained by quantifying the entanglement of one single impurity-spin with the rest of the system. This can be done by calculating the von Neumann entropy $E$ of the reduced density matrix of the impurity spin $\rho_A$ (equivalent results are obtained for the impurity spin $B$). We find 
\begin{eqnarray}
E(\rho_A) &=& -\frac{1+\langle S^z_A + S^z_B\rangle}{2} \log_2 \frac{1+\langle S^z_A + S^z_B\rangle}{2}  \\
&-&\frac{1-\langle S^z_A + S^z_B\rangle}{2} \log_2 \frac{1-\langle S^z_A + S^z_B\rangle}{2},
\nonumber
\end{eqnarray}
where $\rho_A = (\sigma_A^0 + \langle S^z_A + S^z_B\rangle \sigma_A^z)/2$ is independent of the correlation function $\langle {\bf S}_A \cdot {\bf S}_B\rangle$. The $E(\rho_A)$ is bounded between zero (uncorrelated spin-impurity) and one (fully entangled spin-impurity). Full entanglement of single impurity-spins is reached for $\langle S^z_A + S^z_B\rangle=0$, in agreement with previous results for global impurity-spin/electron singlet states.\cite{CMc-PRA06} The presence of a finite $\langle S^z_A + S^z_B\rangle$ decorrelates the single impurity spins from the rest of the system. Full decorrelation (separation) would require $\langle S^z_A + S^z_B\rangle=\pm 1$. However, symmetry constraints impose $\langle S^z_A + S^z_B\rangle \le 2/3$ (see Fig.~\ref{fig-psd}) forcing the entropy $E(\rho_A)$ to be lower-bounded by a finite value close to 0.65. This means that single impurity spins remain highly correlated with the rest of the system even for odd electronic filling.

%
%


\section{Numerical simulations on quantum corrals}
\label{NSQC}

The results discussed above are valid for general two-impurity models interacting with a confined electronic bath. 
We now perform a case study of the entanglement features for 
two impurities located at the foci of an elliptic corral 
confining surface electrons. Each impurity interacts antiferromagnetically with the spin of the band electron located at the corresponding 
focus (Fig. \ref{fig-ks}), modeled by the Hamiltonian:
\begin{equation}
\label{H}
H=H_{\rm e} + J( {\bf S}_A . {\boldsymbol \sigma}_A + {\bf S}_B .{\boldsymbol \sigma}_B ),
\end{equation}
with $J > 0$, 
\begin{equation}
{\bold S}_i . {\boldsymbol \sigma}_i = S_i^z . \sigma_i^z + \frac{1}{2}(S_i^+ .    
\sigma_i^- + S_i^- . \sigma_i^+),
\end{equation}
$\sigma_i^+=c_{i\uparrow}c_{i\downarrow}$, and $\sigma_i^z=
(n_{i\uparrow}-n_{i\downarrow})/2$. Here, $n_{i\sigma}$ and $c_{i\sigma}$ are, 
respectivelly, the number and destruction operators for electrons with spin $\sigma$
on focus $i=A,B$. In the basis of eigenstates $|\alpha\rangle$ of
the ellipse, these local operators can be expanded as $c_{i\sigma}=\sum_{%
\alpha} \Psi_{\alpha i}c_{\alpha \sigma}$ where $c_{\alpha \sigma}$ and $%
\Psi_{\alpha i}$ are the destruction operator and amplitude in state $%
|\alpha\rangle$. In this basis the electronic spin operators are expressed as:
\begin{eqnarray}
\sigma_i^z&=&\frac{1}{2}\sum_{\alpha_1 \alpha_2} \Psi^*_{\alpha_1
i}\Psi_{\alpha_2 i} (c_{\alpha_1 \uparrow}^\dagger c_{\alpha_2 \uparrow} -
c_{\alpha_1 \downarrow}^\dagger c_{\alpha_2 \downarrow}),  \nonumber \\
\sigma_i^+&=&\sum_{\alpha_1 \alpha_2} \Psi^*_{\alpha_1 i}\Psi_{\alpha_2 i} 
c_{\alpha_1 \uparrow}^\dagger c_{\alpha_2 \downarrow}.
\end{eqnarray}
The $H_{\rm e}$ in Eq.~(\ref{H}) accounts for electron confinement in an ellipse   
with hard walls (eccentricity $\epsilon=0.6$ in numerical simulations). \cite{NHA-PRB07}

We perform many-body simulations of Eq. (\ref{H}) applying numerical techniques as 
exact diagonalization or Lanczos, considering up to 10 electronic levels in the ellipse
and different particle fillings $N$ (either even or odd). 
For odd electronic fillings we consider just the $S_{\rm T}^z=1/2$ subspace. Results for $S_{\rm T}^z=-1/2$ are equivalent due to up-down spin symmetry (something reflected by the left-right symmetry of Figs. \ref{fig-psd}, \ref{fig-vb} and \ref{fig-E}). 
We also chose the Fermi energy to coincide with the $23^{\rm rd}$ eigenstate $|\alpha_{23}\rangle $, 
reproducing the experimental conditions of Ref. \onlinecite{MLE-Nature00}.
This configuration is not crucial for the main conclusions: other eccentricities and Fermi 
levels show similar features as long as the wave function bears an
appreciable weight at the foci of the ellipse.

We first follow Eq. (\ref{SzFs}) by calculating numerically the spin correlation function, $\langle {\bf S}_A \cdot {\bf S}_B\rangle$, and the z-projection of the localized spins, $\langle S^z_A + S^z_B\rangle$, for different values of the coupling strength $J$. We find a coefficient $a_4=0$ for small odd fillings ($N \le 3$) and a $a_4 \neq 0$ otherwise, leading to a non-linear relation between the observables. Results for 10 levels and 9 particles are plotted in Fig. \ref{fig-psd} (curve H represented by a broken line). By following its route along the phase-space diagram one observes the entanglement features of the impurity spin subsystem under such conditions, running from separable to entangled areas within the $\langle {\bf S}_A \cdot {\bf S}_B\rangle > 0$ region.
More complex behaviors are to be expected for less restrictive symmetries as shown in Ref. \onlinecite{sarandy}. Results for even fillings are indicated with a broken line coinciding with curve 1.

Finally, we quantify the entanglement developed between the localized impurities and the electronic bath in the ellipse based on Eq.~(\ref{E-AB}). The results are depicted in Fig.~\ref{fig-E}: the broken line for an even filling (10 levels and 10 particles) and the full line for an odd filling (10 levels and 9 particles). Here we notice that the entanglement monogamy\cite{monogamy} is clearly satisfied by observing that a large entanglement entropy between impurity spins and electrons corresponds to a low concurrence between impurity spins, and vice versa.


\section{Conclusions}

We have identified the ground-state entanglement properties of two magnetic impurities (1/2 spins or qubits) interacting with confined electronic backgrounds by composing a phase diagram for the classification of 
separable, entangled, and unphysical states. This was done by studying the impurity-spin concurrence as a 
function of the spin correlation function and the $z$-projection of the localized spins. Such quantum correlations 
show a striking dependence on the electronic filling: while an even filling reduces the impurity spin system to a highly symmetric Werner state, an odd filling breaking spin rotation invariance give rise to richer features. We also discussed the relation between entanglement and (non)locality by studying the conditions for the violation of Bell inequalities in the impurity spin system. Additionally, we obtained complementary information by studying the entanglement developed between the the impurity spins and the electronic background. 

As a case study, we performed corresponding numerical simulations for elliptic quantum corrals consisting of a pair of impurity spins located at the foci coupled antiferromagnetically to a background of surface electrons confined by the elliptic potential. By increasing the interaction parameter, there exists a transition from an RKKY-like regime (where the impurity spins are mutually entangled) to a localized Kondo-like regime (where impurity spins decorrelate, developing a strong entanglement with the electronic environment). We notice that electron confinement may help to preserve quantum correlations by protecting the impurity spin systems from external fluctuations. Moreover, impurity spin correlations could be accessible experimentally by implementing magnetic STM tips or similar devices. We also expect these results to be valid for other kinds of backgrounds such as spin chains. Quantum corrals offer an ideal scenario to study entanglement between impurities and corral 
wave functions or other possible partitions. This quantum information interpretation is an 
alternative way of 
analyzing crossovers between different quantum states and sheds light onto other relevant properties 
stemming from focalization and confinement. 


\acknowledgments
We acknowledge enlightening discussion with Armando Aligia.
We also acknowledge support from the Ram\'on y Cajal program, from the
Spanish Ministry of Science and Innovation's projects No.
FIS2008-05596 and FIS2011-29400, and from the Junta de Andaluc\'ia's Excellence
Project No. P07-FQM-3037.



\begin{thebibliography}{99}


\bibitem{NC-00}
M. Nielsen and I. Chuang, {\it Quantum Computation and Quantum Information} (Cambridge University Press, Cambridge, 2000).

\bibitem{Z-11}
A.M. Zagoskin, {\it Quantum Engineering: Theory and Design of Quantum Coherent Structures} (Cambridge University Press, Cambridge, 2011).

\bibitem{MLE-Nature00}
H.C. Manoharan, C.P. Lutz, and D. Eigler, Nature (London) {\bf 403}, 512 (2000)

\bibitem{FH-RMP03}
G. Fiete and E. Heller, Rev. Mod. Phys. 75, 933 (2003); A. A. Aligia and A. M. Lobos, J. Phys.: Condens. Matter {\bf 17}, S1095 (2005).

\bibitem{HCB-PRL02}
K. Hallberg, A. A. Correa, and C. A. Balseiro, Phys. Rev. Lett. {\bf 88}, 066802 (2002); A. Correa, K. Hallberg, and C.A. Balseiro, EPL {\bf 58}, 6 (2002)

\bibitem{agam} O. Agam and A. Schiller, Phys. Rev. Lett. {\bf 86} , 484 (2001).

\bibitem{aligia} A. A. Aligia, Phys. Rev. B {\bf 64}, 121102 (2001).

\bibitem{porras} D. Porras, J. Fern\'andez-Rossier and C. Tejedor, Phys.
Rev. B {\bf 63}, 155406 (2001).

\bibitem{fiete} G. A. Fiete, J. S. Hersch, E. J. Heller, H. C. Manoharan, C. P. Lutz, and D. M. Eigler,  Phys. Rev. Lett. {\bf 86}, 2392 (2001).

\bibitem{armandochiappe} G. Chiappe and A. A. Aligia, Phys. Rev. B {\bf 66},
075421 (2002).

\bibitem{lobos} A. Lobos and A. A. Aligia, Phys. Rev. B {\bf 68}, 035411 (2003).

\bibitem{kampf} M. Schmid and A. Kampf, Ann. der Physik, {\bf 12}, 463 (2003); ibid. {\bf 14}, 556 (2005).

\bibitem{rossi} E. Rossi and D. K. Morr, Phys. Rev. Lett. {\bf 97}, 236602 (2006).

\bibitem{moon} C. Moon, C. Lutz and H. Manoharan, Nature Physics {\bf 4}, 454 (2008).

\bibitem{NHA-PRB07}
M. Nizama, K. Hallberg, and J. d'Albuquerque e Castro, Phys. Rev. B {\bf 75}, 235445 (2007).

\bibitem{NFH-PB09}
M. Nizama, D. Frustaglia, and K. Hallberg, Physica B {\bf 404}, 2819 (2009).

\bibitem{PTP} K. Hallberg and M. Nizama, Prog. Theor. Phys. {\bf 176}, 408 (2008). 

\bibitem{WB-PE01} M . Weissmann and H. Bonadeo, Physica E (Amsterdam) {\bf 10}, 544 (2001).

\bibitem{S-PRL06} J. Simonin, Phys. Rev. Lett. {\bf 97}, 266804 (2006). 

\bibitem{MJ-PRB09} D.F. Mross and H. Johannesson, Phys. Rev. B {\bf 80}, 155302 (2009).

\bibitem{W-Science01}
S.A. Wolf {\it et al.}, Science {\bf 294}, 1488 (2001); B. E. Kane, Nature (London) {\bf 393}, 133 (1998).

\bibitem{MS-PRL04}
D. K. Morr and N. A. Stavropoulos, Phys. Rev. Lett. {\bf 92}, 107006 (2004); Z.-G. Fu, P. Zhang, Z. Wang, and S.-S. Li, Phys. Rev. B {\bf 84}, 235438 (2011).

\bibitem{RKKY}
M. A. Ruderman and C. Kittel, Phys. Rev. {\bf 96}, 99 (1954); T. Kasuya, Prog. Theor. Phys. {\bf 16}, 45 (1956); K. Yosida, Phys. Rev. {\bf 106}, 893 (1957).

\bibitem{note-2}
The hybridization with bulk states would produce a broadening of the discrete levels in the corral. 
Our results are valid as long as the levels involved do not overlap (see Ref.~\onlinecite{HCB-PRL02}). 
For large hybridizations, the total number of particles and total spin projection are not good quantum 
numbers causing other kinds of entanglement.

\bibitem{CMc-PRA06}
S.Y. Cho and R.H. McKenzie, Phys. Rev. A {\bf 73}, 012109 (2006).

\bibitem{JVW-PRL88}
B.A. Jones, C.M. Varma, and J.W. Wilkins, Phys. Rev. Lett. {\bf 61}, 125 (1988); {\it ibid.} {\bf 61}, 2819 (1998); B.A. Jones and C.M. Varma, Phys. Rev. B {\bf 40}, 324 (1989).

\bibitem{W-PRA89}
R.F. Werner, Phys. Rev. A {\bf 40}, 4277 (1989).

\bibitem{W-PRL98}
W.K. Wootters, Phys. Rev. Lett. {\bf 80}, 2245 (1998).

\bibitem{note} This is straightforward considering rotational invariance for the singlet and  up-down spin symmetry for the doublet.

\bibitem{note-1}
This can be alternatively obtained directly from the density matrix by noticing that its diagonal elements must be positive.

\bibitem{CHSH-PRL69}
J.F. Clauser, M.A. Horne, A. Shimony, and R.A. Holt, 
Phys. Rev. Lett. {\bf 23}, 880 (1969). 

\bibitem{HHH-PLA95}
R. Horodecki, P. Horodecki, and M. Horodecki,
Phys. Lett. A {\bf 200}, 340 (1995).

\bibitem{BBPS-PRA96}
C.H. Bennett, H.J. Bernstein, S. Popescu, and B. Schumacher, 
Phys. Rev. A {\bf 53}, 2046 (1996).

\bibitem{sarandy} 
M. S. Sarandy, Phys. Rev. A {\bf 80}, 022108 (2009).

\bibitem{monogamy}
V. Coffman, J. Kundu, and W. K. Wootters, Phys. Rev. A {\bf 61}, 052306 (2000); 
B. M. Terhal, IBM J. Res. Dev. {\bf 48}, 71 (2004); 
M. Koashi and A. Winter, Phys. Rev. A {\bf 69}, 022309 (2004). 



\end{thebibliography}
\end{document}